\input{aipcheck}
\documentclass[
    ,final            
  ]
{aipproc}

\layoutstyle{6x9}

\begin{document}

\title{A novel background reduction strategy for high level triggers and processing in gamma-ray Cherenkov detectors}


\author{G.~Cabras$^{*}$, A.~De~Angelis$^{*}$, B.~De~Lotto$^{*}$, M.M.~De~Maria$^{*}$, F.~De~Sabata$^{*}$, O.~Mansutti}
{address={Dipartimento di Fisica dell'Universit\`a di Udine and INFN, Udine, Italy}
 }
\author{M.~Frailis$^{\dag}$, M.~Persic}
{address={Osservatorio Astronomico di Trieste and Dipartimento di Fisica dell'Universit\`a di Udine, Italy }
} 
\author{C.~Bigongiari$^{**}$, M.~Doro$^{**}$, M.~Mariotti$^{**}$, L.~Peruzzo$^{**}$, A.~Saggion$^{**}$, V.~Scalzotto}
{address={Dipartimento di Fisica dell'Universit\`a di Padova and INFN, Padova, Italy}
}
\author{R.~Paoletti$^{\ddag}$, A.~Scribano$^{\ddag}$, N.~Turini}
{address={Dipartimento di Fisica dell'Universit\`a di Siena and INFN, Siena, Italy}
}
\author{A.~Moralejo$^{\S}$, D.~Tescaro}
{address={Institut de F\`isica d'Altes Energies, Edifici Cn., E-08193 Bellaterra (Barcelona), Spain}
}
\author{the MAGIC Collaboration}{address={http://wwwmagic.mppmu.mpg.de/}}

\begin{abstract}
Gamma ray astronomy is now at the leading edge for studies related both to fundamental physics and astrophysics.
The sensitivity of gamma detectors is limited by the huge amount of background, constituted by hadronic cosmic rays (typically two to three orders of magnitude more than the signal) and by the accidental background in the detectors.
By using the information on the temporal evolution of the Cherenkov light, the background can be reduced.
We will present here  the results obtained within the MAGIC experiment using a new technique for the reduction of the background.
Particle showers produced by gamma rays show a different temporal distribution with respect to showers produced by hadrons; the background due to accidental counts shows no dependence on time.
Such novel strategy can increase the sensitivity of present instruments. 
\end{abstract}

\maketitle



During the recent years, there has been an impressive progress in gamma-ray astrophysics thanks to ground-based gamma 
detectors. Such a progress has impact both on fundamental physics and on astrophysics \cite{deareview}.

Ground-based telescopes cannot detect cosmic gamma rays directly, as such particles are absorbed when entering the atmosphere.
They detect instead the radiation emitted by the ultra-relativistic particles of the showers produced by the interaction of photons with the atmosphere.
In particular, IACTs \emph{(Imaging Atmospheric Cherenkov Telescopes)} detect the Cherenkov light emitted by secondary charged particles of a shower, and study the properties of the primary particle from the image formed by the Cherenkov light on photodetectors placed in the focal plane.
The physical and phenomenological characterization of these showers, and the morphological study of the shape of the Cherenkov image collected in the camera plane, give the possibility to select the showers induced by the gamma events among the huge amount of  cosmics.

MAGIC is a IACT located in the Canary island of La Palma ($28.75^\circ$N, $17.86^\circ$W, 2225~m a.s.l.); its  reflecting surface, 17~m~in diameter, has a parabolic shape (which preserves the time structure of the Cherenkov light flash).
The field of view of the focal PMT camera is about $3.5^\circ$.
Fast PMT analog signals are routed via optical fibers to the DAQ-system electronics, where 
the signals are digitized and saved to disk. 
Further details can be found in \cite{factsheet}.

Thanks to a major upgrade, since February 2007 the data acquisition
works via ultra-fast FADCs, digitizing the signal 
at the speed of 2~GSamples/s \cite{2007ICRCGoebel}.
Particle showers produced by gamma rays have a temporal distribution
of $(2\div5)$ ns, narrower than the ones typical for hadronic showers, of around
$(5\div10)$~ns.
Using the time structure of the signal, we can obtain an enhancement of the telescope performance basically for two reasons: a reduction in the amount of NSB (Night Sky Background) light integrated with the real signal (due to a smaller integration window), and the possibility to reconstruct with a good time resolution the timing characteristics of the showers \cite{Tescaro} \cite{tesi1,tesi2}.
 

For this work, to find the maximum pulse within the useful range of the FADCs
($\sim30$~ns), we use a cubic spline function. The position of its
maximum is used to estimate the arrival time of the signal in a pixel.
The intensity of the signal is then obtained by integrating the spline in a range of $7.5$~ns.
In this way, the time resolution of each pixel has been estimated to
be around $0.4$~ns~RMS for a signal of 40 photoelectrons $(phe)$, through the study of the
pixel time spread in calibration events. This value may improve with the use of a more sophisticated pulse reconstruction.

\begin{figure}[h!]
\includegraphics[width=0.9\textwidth]{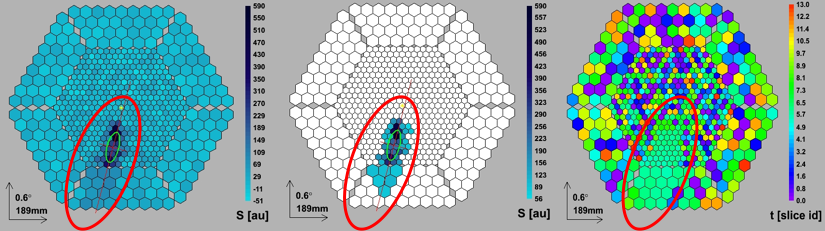}
\caption{Shower image
  before (left) and after (center)image cleaning. The ellipse superimposed on the
  camera corresponds to the Hillas approximation. The third map (right) shows
  the arrival time for each pixel: it can be seen that there is a
  tight coincidence for pixels belonging to the gamma shower image.} 
\label{img:Cleaning}
\end{figure}

After the calibration of the data, for each event, the shower image appears like in figure
\ref{img:Cleaning}, on the left: it is surrounded by pixels containing
only noise from the light of the diffuse background.
The image cleaning procedure removes pixels which apparently do not form part of
the shower image (figure \ref{img:Cleaning}, center). 

Only low energy events have been selected, through an upper cut in the signal of the two highest pixels, since we want to
concentrate our algorithm on the region of the order of 50 GeV, the most difficult to detect for Cherenkov telescopes.
We require a time coincidence between core and
boundary pixels within a window spanning from $0.5$ to $3.5$~ns.
After the cleaning, the shape of the surviving image is studied by an approximation to an ellipse, as introduced by Hillas \cite{1985ICRC....3..445H}. The information is therefore saved into different
parameters related with this ellipse, like its width, its length and
its orientation in the camera.
To test the performance of the cleaning, we study the enhancement of the signal to noise ratio obtained after a cut in
the Hillas ALPHA parameter\footnote{ALPHA indicates the orientation of the
image ellipse with respect to the source position: for gammas it
is expected to be close to 0.}.

\begin{figure}[t]
\includegraphics[width=0.9\textwidth]{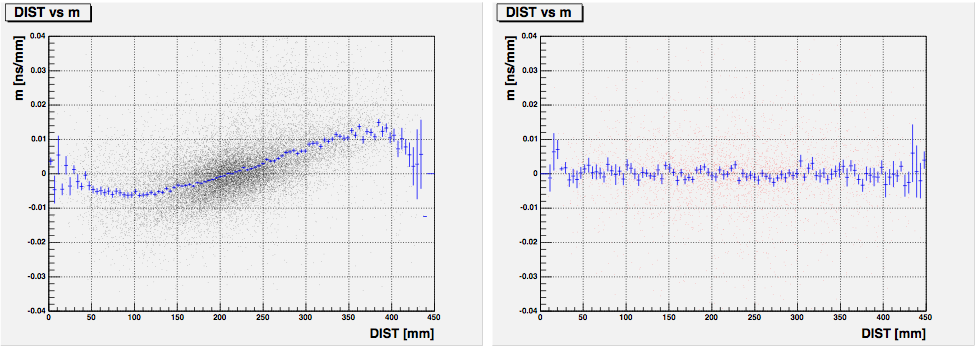}
\caption{TIME GRADIENT (called $m$) vs DIST ($297$~mm$=1~deg$). The
correlation for simulated $\gamma$-rays is evident (left); this correlation does
not appear for hadrons (right).} \label{fig:time_gradient}
\end{figure}

The information given by the time evolution of the images, has
moreover been used for developing new parameters in the gamma/hadron separation algorithm based on a particular neural network architecture called
Random Forest \cite{randomfor}. We studied the additional background
rejection power added to the classical Hillas parameters.

The TIME~RMS parameter estimates the arrival time spread of the Cherenkov photons in the pixels belonging to the cleaned image ($\mu$,  $\gamma$-ray and hadron induced showers  have a different characteristic time spread~\cite{2006APh....25..342}).
Previous studies~\cite{1999APh....11..363H} determined that along the major axis of gamma induced Cherenkov image a time structure is present (see figure 1, right). This can be well approximated as a linear TIME~GRADIENT.
The slope of this gradient is related to the angle between the telescope
and shower axes; moreover, for showers of a given direction, it is
well correlated with the Impact Parameter (IP) of the shower, and so
with the DIST image parameter\footnote{IP: distance between the telescope
pointing axis and the axis of the shower. In case of real data
DIST is a good estimator of the IP.}, as can be seen in figure \ref{fig:time_gradient}.

A data sample of 5.6 h of Crab Nebula observation (February 2007; zenith angle between $5^\circ$ and $30^\circ$) 
has been used to check the results on real data; the results are consistent with the ones previously described.

\begin{figure}[t]
\includegraphics[width=0.9\textwidth]{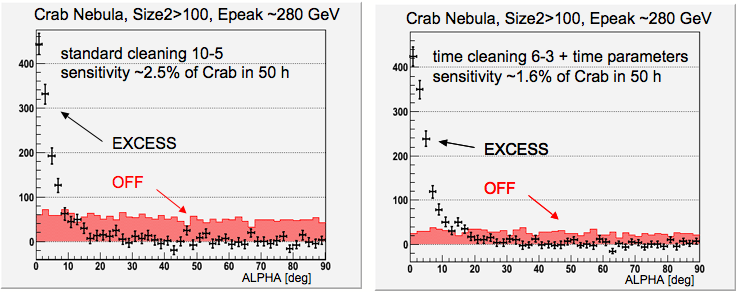}
\caption{Comparison of two $\alpha$-plots between analysis 1 (left) and 3 (right). Keeping the same number of excess events background is reduced by about 50\%. Energy threshold: 280~GeV.} \label{fig:alpha}
\end{figure}

Three different analyses of the same previously described Crab~Nebula
data have been performed: the first one by a standard analysis, the
second one by an image cleaning (IC) with time constraint, and the third one by a
time-IC and the TIME RMS and TIME GRADIENT in the Random Forest
parameter list. In figure \ref{fig:alpha} we compare two $\alpha$-plots obtained from
the first and the third analysis.

Time parameters allows $\sim$50\% better background suppression keeping the same amount of excess events. Similar improvements are seen at all energies.
Improvements (of ~15\%) have been found also in the event energy reconstruction. In fact the TIME GRADIENT gives information about the real IP of the shower and therefore it helps to distinguish distant high energy showers from closer, low energy ones.

We can conclude that the use of timing parameters in the analysis of
current MAGIC data results in a significant improvement in background
rejection, and thus in sensitivity. 

The integral sensitivity of the telescope obtained with the standard analysis for energies
larger than 280 GeV was 2.5\% of Crab in 50 hours, while through this novel
analysis we obtain a value of 1.6\% of Crab.

Moreover, at the lowest energies the time 
cleaning also contributes, besides reducing the background, to enlarge the event statistics.

The new scenario for temporal analysis of Cherenkov flashes has just
been explored for the first time, and  further developments are
close to come.
Within the year 2008, MAGIC will be upgraded to a two telescope system. A dedicated Monte Carlo chain has been developed in order to estimate its performance: observations in stereoscopic mode will increase
the current sensitivity of the instrument by a factor ~3 \cite{MonteCarlo}.

\begin{theacknowledgments}
This work was partly financed by the Ministero dell'Universit\`a e della Ricerca within the PRIN 2005.
\end{theacknowledgments}

\end{document}